\documentclass[floatfix,intlimits,showkeys,twocolumn,superscriptaddress]{revtex4-1}

\usepackage{amsfonts,amsmath,amssymb}
\usepackage{dcolumn}
\usepackage{graphicx}
\usepackage[utf8x]{inputenc}
\usepackage{textcomp}
\usepackage{textgreek}
\usepackage{todonotes}
\usepackage{hyperref}

\presetkeys{todonotes}{inline}{}

\newcommand{\degree}{\ensuremath{^\circ}}%
\newcommand{\eg}{e.\,g.}%
\newcommand{\ie}{i.\,e.}%

\begin{document}

\title{Alternating-Gradient Focusing of the Benzonitrile-Argon Van der Waals Complex}%

\author{Stephan Putzke}%
\author{Frank Filsinger}%
\affiliation{Fritz-Haber-Institut der Max-Planck-Gesellschaft, Faradayweg 4-6, 14195 Berlin, Germany}%
\author{Jochen K\"upper}%
\affiliation{Center for Free-Electron Laser Science, DESY, Notkestrasse 85, 22607 Hamburg, Germany}%
\affiliation{University of Hamburg, Department of Physics, Luruper Chaussee 149, 22761 Hamburg, Germany}%
\author{Gerard Meijer}%
\email[Electronic mail: ]{meijer@fhi-berlin.mpg.de}%
\affiliation{Fritz-Haber-Institut der Max-Planck-Gesellschaft, Faradayweg 4-6, 14195 Berlin, Germany}%

\date{\today}%

\begin{abstract}\noindent%
We report on the focusing and guiding of the van der Waals complex formed between benzonitrile molecules (C$_6$H$_5$CN) and argon atoms in a cold molecular beam using an ac electric quadrupole guide. The distribution of quantum states in the guided beam is non-thermal, because the transmission efficiency depends on the state-dependent effective dipole moment in the applied electric fields. At a specific ac frequency, however, the excitation spectrum can be described by a thermal distribution at a rotational temperature of 0.8~K. From the observed transmission characteristics and a combination of trajectory and Stark-energy calculations we conclude that the permanent electric dipole moment of benzonitrile remains unchanged upon the attachment of the argon atom to within $\pm$5\%. By exploiting the different dipole-moment-to-mass (\textmu/m) ratios of the complex and the benzonitrile monomer, transmission can be selectively suppressed for or, in the limit of 0~K rotational temperature, restricted to the complex.
\end{abstract}%
\maketitle%

\section{Introduction}
\label{sec:intro}
The motion of neutral polar molecules can be controlled by electric fields. However, the manipulation of neutral molecules is
challenging, because their interaction with an electric field, through the quantum mechanical Stark effect, is several
orders of magnitude weaker than the electrostatic interaction with ions. Many techniques that have been developed over
the last decades resemble methods used for the manipulation of charged particles. Examples include the focusing,
acceleration/deceleration, and trapping of polar molecules using switched and static electric fields~\cite{Meerakker:NatPhys4:595,Wall:PRA80:043407,Hogan:2011fk,Meerakker:2012uq} (and references therein). In this work we report on a dipole-moment-to-mass selector for polar molecules, the equivalent of a quadrupole mass filter (QMF) for ions. Neutral molecules are selectively guided or deflected out of a molecular beam depending on their dipole-moment-to-mass
(\textmu/m) ratio, analog to the charge-to-mass (q/m) ratio selectivity for ions by the QMF.

\begin{figure}[t]
   \centering
   \includegraphics[width=\linewidth]{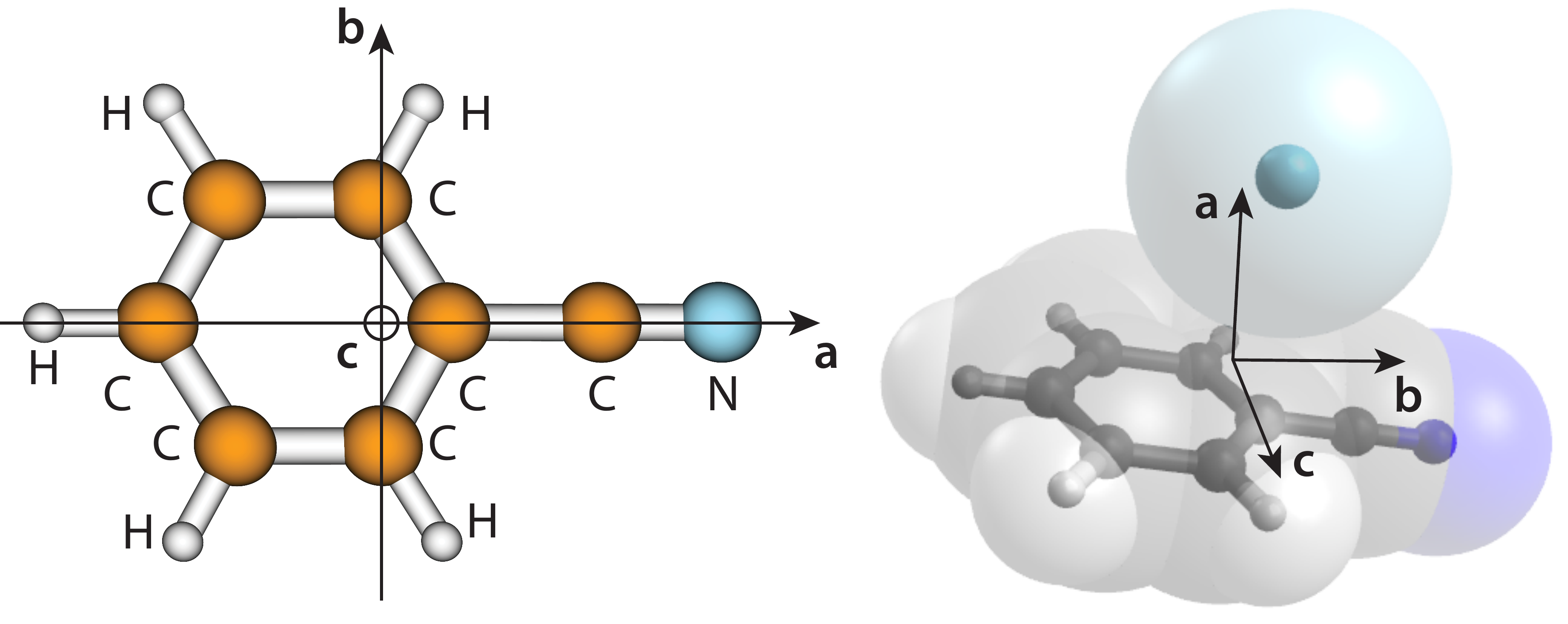}
   \caption{Structure of benzonitrile (left) and the benzonitrile-argon complex (right)~\cite
{Helm:CPL270:285}. The principal axes are indicated and van-der-Waals spheres are drawn around the atoms of the complex as an illustration.}
   \label{fig:molecules}
\end{figure}

Molecules in polar quantum states are either attracted towards regions of low or high electric
fields, effectively dividing them into two classes, so-called low-field-seeking (lfs) and high-field-seeking (hfs)
states. Small molecules, such as CO, NH$_3$ and OH, have both lfs and
hfs states, and are often manipulated in their lfs states by static electric fields. However, large molecules, such as
benzonitrile, have only hfs states at already small electric field strengths. In this case, the use of static electric
fields to confine the molecules to the molecular beam axis, for instance, is not possible as no electric field
maximum can be created in free space. Dynamic focusing schemes, \ie, alternating-gradient (AG) focusing,
need to be applied instead~\cite{Auerbach:JCP45:2160}. AG focusing is implemented in our focuser by alternating
between two configurations of the electric field, thereby guiding the molecules on stable
trajectories through the device. Using an AG focuser, we have demonstrated the selective transmission of
individual conformers of large molecules~\cite{Filsinger:PRL100:133003}, as well as the
change of the forward velocity of large molecules~\cite{Wohlfart:PRA77:031404}. More recently, at an increased
\textmu/m-resolution, we restricted the transmission to molecules in their rovibronic ground state, effectively cleaning the
molecular beam from molecules in higher rotational states that are still populated even at the low rotational temperature of
about 1~K in the beam~\cite{Putzke:2011fk}. In all these experiments the transmission of molecular species
with the same mass but different dipole moments was studied. In this work, we compare the transmission of benzonitrile (103~u)
and the benzonitrile-argon van der Waals complex (143~u), both visualized in \autoref{fig:molecules}, in their electronic,
vibrational and rotational ground states. The permanent electric dipole moment of the
benzonitrile monomer is expected not to change significantly upon the attachment of the argon atom, because the parent molecular
structure remains unchanged~\cite{Helm:CPL270:285}. Therefore, transmission can be studied at (nearly) identical dipole
moments but at different masses.

\section{Experimental Setup}
\label{sec:setup}

A pulsed molecular beam containing the benzonitrile-argon complex is produced by bubbling argon gas at 4~bar through
liquid benzonitrile at 55\degree{C}. The mixture is then expanded into vacuum through a heated (60\degree{C}) pulsed
valve with an 0.8~mm diameter orifice (General Valve, Series 99) at a 40~Hz repetition rate and the resulting forward velocity of the beam is about 600~m/s. After passing through a 1.5~mm diameter skimmer placed 40~mm downstream from the nozzle, the molecules enter the guide 32~mm behind the tip of the skimmer. The guide, described in more detail in reference~\citealp{Putzke:2011fk}, consists of four 1.64~m long cylindrical stainless steel electrodes with a diameter of 6~mm and a closest surface-to-surface distance between adjacent electrodes of 1.42~mm. Special care is taken to keep this distance constant within $\pm$0.02~mm over the full length of the device.

High voltages ($\pm$9.5~kV) are applied to the electrodes, creating a 48~kV/cm saddle point of the electric field on the molecular beam axis. This results in an acceleration of high-field-seeking molecules toward the molecular beam axis in one direction and an acceleration away from the axis in the perpendicular direction. The roles of the two axes can be interchanged by interchanging the voltages on the electrodes, effectively rotating the electric field by 90\degree. By rapidly (rise-times \textless0.5~\textmu s) switching back and forth between these two electric field configurations at a suitable frequency, alternating-gradient (AG) focusing is achieved, and the molecules are transported through the selector on stable trajectories. In all presented experiments, the two configurations are applied for equal durations in each switching cycle.

The molecules are detected 40~mm downstream from the selector via laser-induced fluorescence (LIF). The 514~nm light of a
thin-disk laser (ELS MonoDisk, operated at 5~W) pumps a continuous-wave ring dye laser (Coherent 899-21, Rhodamine~110). The dye-laser
output is frequency doubled in an external cavity (Spectra Physics, LAS WaveTrain). Typically 20~mW of 274~nm radiation is
obtained, with an instantaneous bandwidth of less than 1~MHz. Stabilizing the laser frequency with respect to a frequency stabilized
helium-neon laser keeps the frequency constant to within 2~MHz for several hours.

The laser beam intersects the molecular beam at 90\degree{} and fluorescence light is collected perpendicular to the plane of laser and molecular beam using a lens system and detected with a photomultiplier tube (PMT). The transmission of the molecules through the guide in individual rotational $J_{K_aK_c}$ levels is measured by counting all detected LIF photons from the entire molecular beam pulse, \ie, time-integrated over the whole pulse.

\section{LIF spectra and discussion}
\label{sec:results}

\begin{figure}
   \centering
   \includegraphics[width=\linewidth]{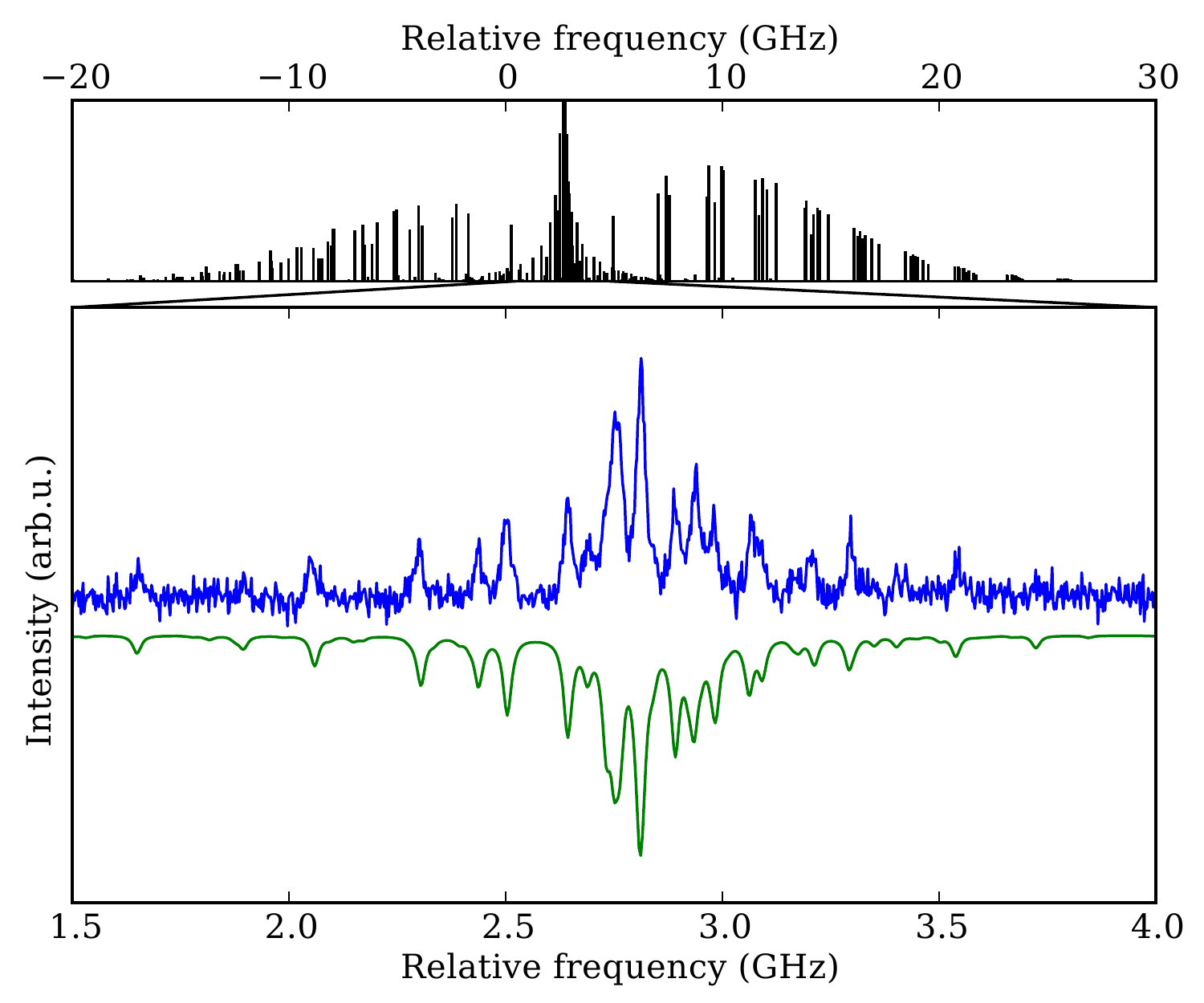}
   \caption{Upper panel: simulation of the rotationally resolved fluorescence excitation spectrum of the complex at a
      rotational temperature of 0.8~K. The $ac$-hybrid spectrum has 93\% $c$-type and 7\% $a$-type contributions. Lower panel: central part of the experimental spectrum obtained at 2.8~kHz switching frequency, and, plotted upside down, simulated spectrum convoluted with a Voigt profile. See text for details.}
   \label{fig:results:spectrum}
\end{figure}

\autoref{fig:results:spectrum} shows part of the rotationally resolved $S_1\leftarrow{}S_0$ electronic fluorescence excitation spectrum of the benzonitrile-argon complex. The experimental spectrum in the lower panel is obtained with the selector operating at 2.8~kHz. The UV-laser frequency is scanned in steps of 2~MHz near the band origin of the electronic transition at 36489.16~cm$^{-1}$. The recorded frequency range is chosen to include the strongest rotational transitions in the Q-branch. At each step, the detected LIF photons from 80 consecutive molecular beam pulses are integrated (2~s). Despite the high resolution of the laser system, not all rotational transitions can be resolved in the Q-branch because of the high density of rotational lines. A simulation of the spectrum using the known molecular constants~\cite{Dahmen:BunsenBer98:970, Meerts:CanJC82:804} with the PGopher program~\cite{Western:pgopher} is shown in the upper panel and a 0.8~K rotational temperature, an adjustable parameter in the simulation, is used. In the lower panel, the simulated spectrum is shown, convoluted with a Voigt profile. We assume that the natural line width of the transition is the same as for benzonitrile which results in a Lorentzian contribution of 8~MHz (full-width-at-half maximum, FWHM) to the line shape. A Gaussian contribution of 21~MHz (FWHM) is obtained from a fit to the experimental spectrum. This Gaussian contribution is mostly due to a 15~MHz Doppler broadening (FWHM) resulting from the transverse velocity distribution of the molecules leaving the focuser, which is 4~m/s (FWHM) according to trajectory simulations. In order to make the effect of transverse forces near the end of the device on the LIF detection similar in all measurements presented in this work, the applied switching sequences are arranged such that molecules in the middle of a pulse leave the selector, when the switching cycle is in the middle of a vertical focusing phase~\cite{Putzke:2011fk}.

\begin{figure}
   \centering
   \includegraphics[width=\linewidth]{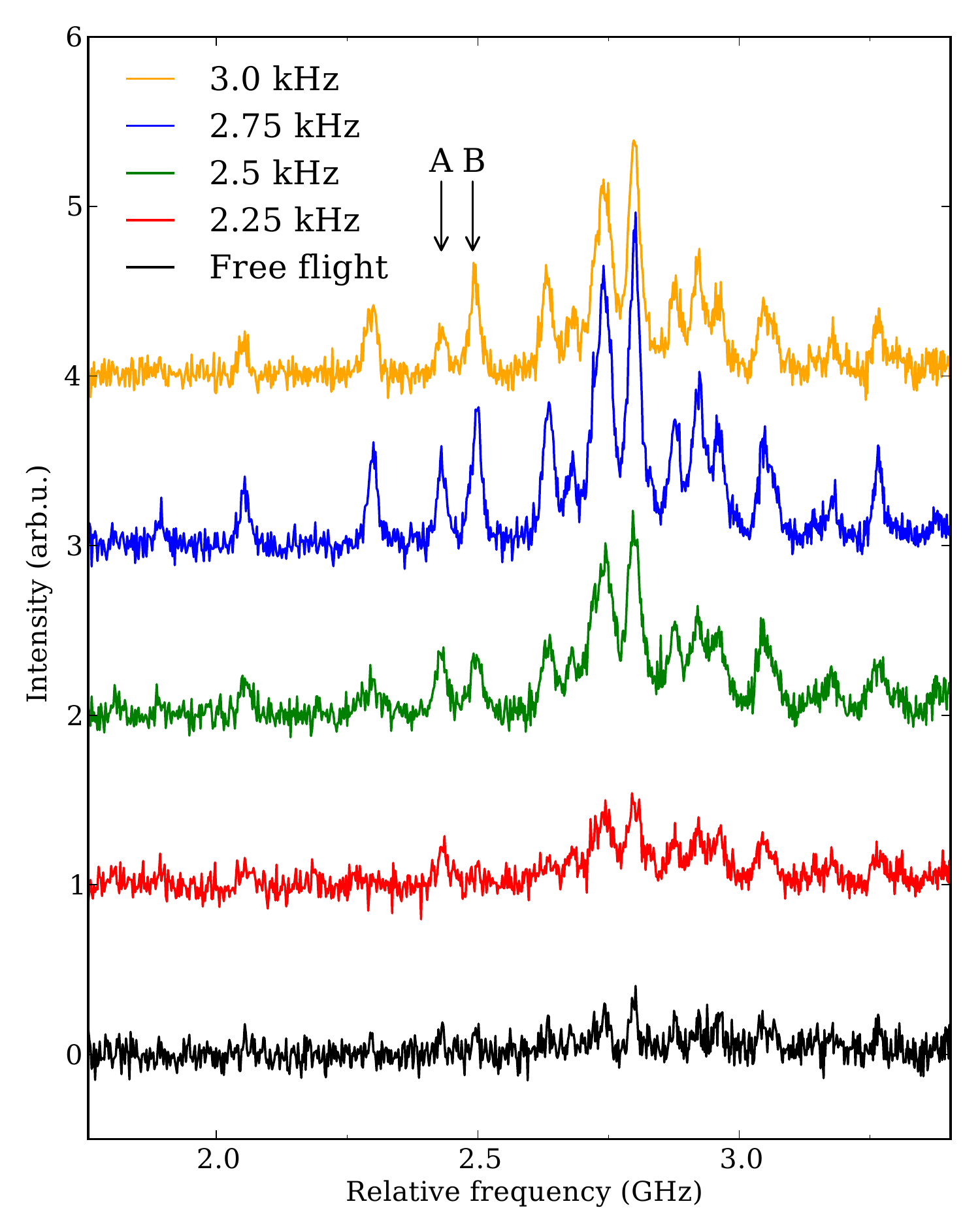}
   \caption{Rotationally resolved fluorescence excitation spectra of benzonitrile-argon (offset for clarity). The black
      bottom spectrum has been obtained at free flight, \ie, without high voltages applied to the selector. The other
      spectra have been obtained with the selector operated at the indicated ac switching frequencies. The laser
      frequency is given relative to the $S_1\leftarrow S_0$ vibronic origin transition at 36489.16~cm$^{-1}$.
      Transitions that contribute to the indicated peaks: $4_{13}\leftarrow 4_{23}$ and $6_{24}\leftarrow 6_{34}$ (A) (ratio 3:1) and $2_{02}\leftarrow 2_{12}$ (B).}
   \label{fig:results:guidingspectrum}
\end{figure}

The peak intensities in the spectrum depend on the ac frequency as can be seen from the spectra measured for different ac frequencies shown in \autoref{fig:results:guidingspectrum}. A weak LIF signal is obtained in free flight (black bottom trace), \ie, when ground potential is applied to the electrodes of the selector and only the strongest transitions (near 2.75~GHz) are clearly observed. Here, the Doppler broadening contributes less than 4~MHz to the line width, because the maximum transverse velocity for molecules that reach the detection region is limited by the geometry of the experimental setup~\cite{Putzke:2011fk}. When the molecules are guided through the selector by applying switched high voltages to the electrodes, the signal is significantly increased. In the spectra obtained with ac frequencies in the 2.25--3.0~kHz range, several distinct peaks are seen and the overall strongest signal is observed at 2.75~kHz. Two peaks in the spectra, labeled (A,B), assigned to the $4_{13}\leftarrow 4_{23}$ and $6_{24}\leftarrow 6_{34}$ (peak A, contributing to the line strength at a 3:1 ratio) and $2_{02}\leftarrow 2_{12}$ (B) transitions, illustrate the dependence of the peak height on the ac frequency. At 2.25~kHz A is strong while B is completely suppressed. When increasing the ac frequency to 3.0~kHz, B gains in intensity faster than A until B is the strongest. It is evident from this behavior that molecules in the initial quantum levels contributing to A have lower effective dipole moments than states that contribute to B (see \autoref{sec:transmissioncurves}). As a result of the guiding process, the contributions from individual $M$-components of the $J_{K_aK_c}$ states to the observed transitional lines are changed in the guided beam compared to the free flight. In this regard it is remarkable that the experimental spectrum is still described well by the (free-flight) simulation at 0.8~K, albeit that the role of the rotational temperature is reduced to an effective parameter.

\begin{figure}
   \centering
   \includegraphics[width=\linewidth]{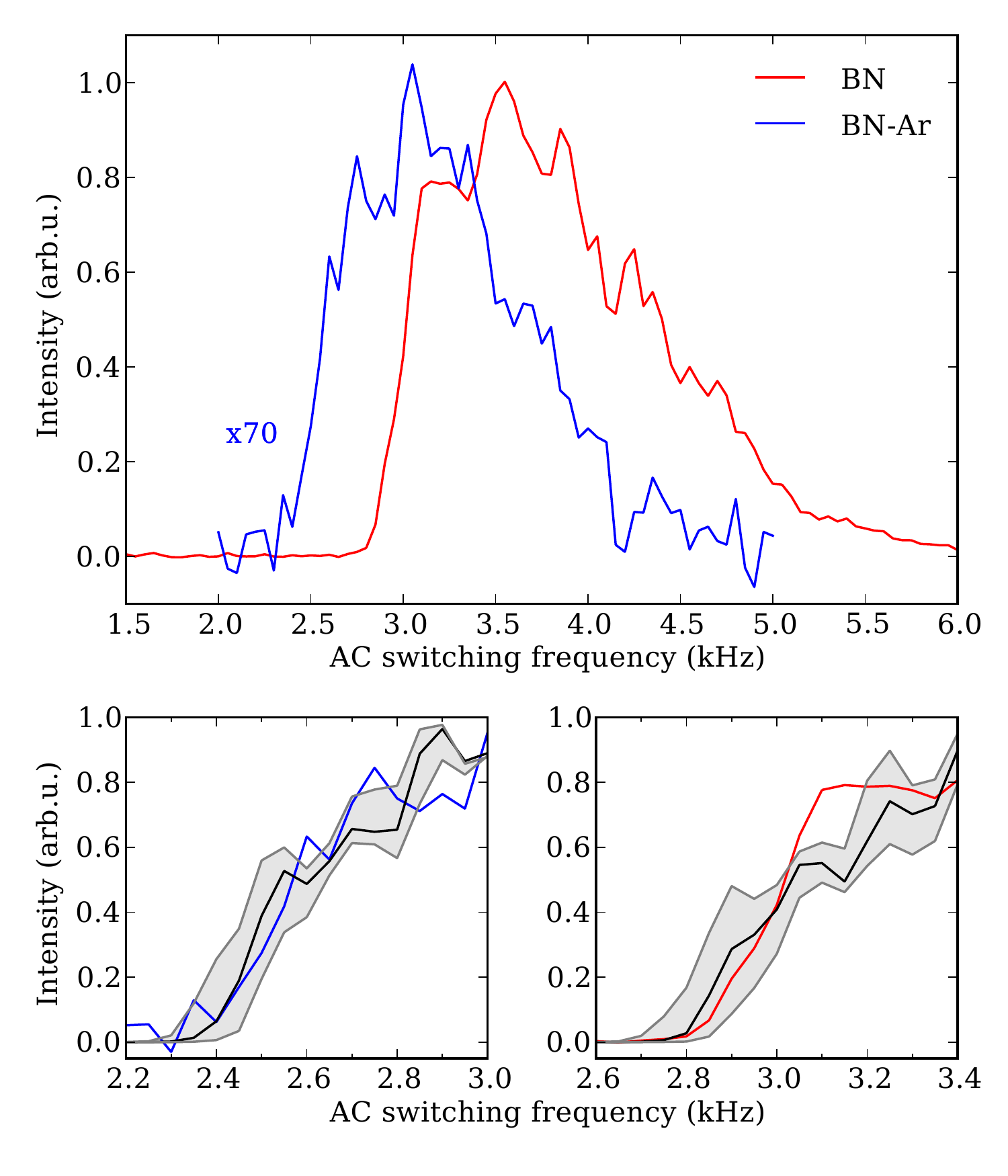}
   \caption{Time-integrated LIF signal intensity as function of ac switching frequency for benzonitrile (red curves) and
      the argon complex (blue curves). The laser frequency is kept fixed at the center of the
      $1_{11}\leftarrow0_{00}$ ($1_{10}\leftarrow0_{00}$) transitions of benzonitrile (the complex) and the complex-signal is scaled 70-fold for better comparison. Lower panels:
      onsets of the curves in the upper panel, plotted on an expanded scale, together with calculated curves (black) for
      molecules with dipole moments of $\mu=4.52$~D and $\mu\pm5\%$ (grey curves). }
   \label{fig:results:transmissioncurves_d05}
\end{figure}

\section{Frequency-dependent transmission curves and discussion}
\label{sec:transmissioncurves}

The transmission efficiency for a $J_{K_aK_c}M$ quantum state in the guiding process depends on the ac frequency, similar to the transmission of ions in the quadrupole mass filter. This dependence is depicted in \autoref{fig:results:transmissioncurves_d05}, where the measured LIF intensity is shown as a function of ac switching frequency for benzonitrile and the complex in their respective rotational ground states. To obtain these transmission curves, the laser frequency is kept fixed at the center of the $1_{11}\leftarrow0_{00}$ (monomer) and $1_{10}\leftarrow0_{00}$ (complex) transitions. The ac frequency is scanned in steps of 50~Hz in the 1.5--6~kHz range and the detected signal from 80 consecutive molecular pulses is integrated. For better comparison, the signal from the complex is scaled up 70-fold.

The relative density of benzonitrile and the complex in the molecular beam, \textrho$_\textrm{BN}$/\textrho$_\textrm{BN-Ar}$, can be estimated from this scaling factor under the following assumptions. The guiding efficiencies for the ground states of the two species are similar and the 1.8~K rotational temperature seen for benzonitrile for free flight is the same for the complex, resulting in a fractional population of the ground state that is 4.1 times higher for benzonitrile than for the complex. Both considered rovibronic transitions have equal line strengths, because there is only one allowed transition from each of the rotational ground states of the respective initial $S_0$,v=0 levels. As a result, the relative density is estimated to: \textrho$_\textrm{BN}$/\textrho$_\textrm{BN-Ar} \approx$ 70 / 4.1=17.

It is seen from the obtained curves in \autoref{fig:results:transmissioncurves_d05} that transmission of molecules is suppressed at low ac frequencies, and, in going to higher frequencies, a steep onset occurs that is followed by a gradual decrease in intensity. The position of the onset indicates the lowest ac frequency at which the AG principle works for molecules in the corresponding quantum state, leading to stable trajectories that extend to the end of the selector. The position critically depends on the geometry and strength of the electric fields in the selector as well as on the effective dipole moment of the molecular quantum state and on the mass. In the lower two panels of \autoref{fig:results:transmissioncurves_d05}, the onsets of the transmission curves from the upper panel are shown on an expanded horizontal scale, and the outcome of trajectory simulations are depicted as black traces. The simulations are based on the experimental switching sequences and on Stark-energy calculations based on the molecular constants~\cite{Wohlfart:JMolSpec247:119,Dahmen:BunsenBer98:970} using the \emph{libcoldmol} program~\cite{Kuepper:libcoldmol}. In the calculations, the permanent electric dipole moment of benzonitrile, oriented along the monomer $a$-axis, is taken for both species: \textmu$_\textrm{BN}$=\textmu$_\textrm{BN-Ar}$=4.52~D. We assume that the dipole moment in the complex is still aligned along the symmetry axis of the benzonitrile monomer, but due to the changed mass distribution in the complex, $\vec{\textrm{\textmu}}_\textrm{BN-Ar}$ is not aligned along a principal axis anymore. The resulting 13\degree{} tilt of the dipole moment relative to the $b$-axis away from the argon atom is included in the calculation and all rotational states with $J\leqslant25$ ($J\leqslant30$ for the complex) are considered.

For both benzonitrile and the complex, the shape of the onset is not perfectly reproduced by the trajectory simulations, but the simulated and measured cut-off frequencies agree very well. In order to study the dependence of the onset position on the magnitude of the dipole moment, the trajectory simulations are repeated with Stark energies calculated from \textmu$_\textrm{BN-Ar}$=\textmu$_\textrm{BN}$ changed by $\pm5\%$. Grey traces depict the resulting simulated transmission curves and the area between these curves is shaded in grey. Simulations in which the tilt of the dipole moment is varied by $\pm$13\degree{} (not shown) do not exhibit such shift of the onset position. From the shifts of the onset position that occur when changing the value and orientation of the dipole moment, we conclude that the permanent dipole moment is the same for benzonitrile and the complex to within $\pm5\%$.

\begin{figure}
   \centering
   \includegraphics[width=\linewidth]{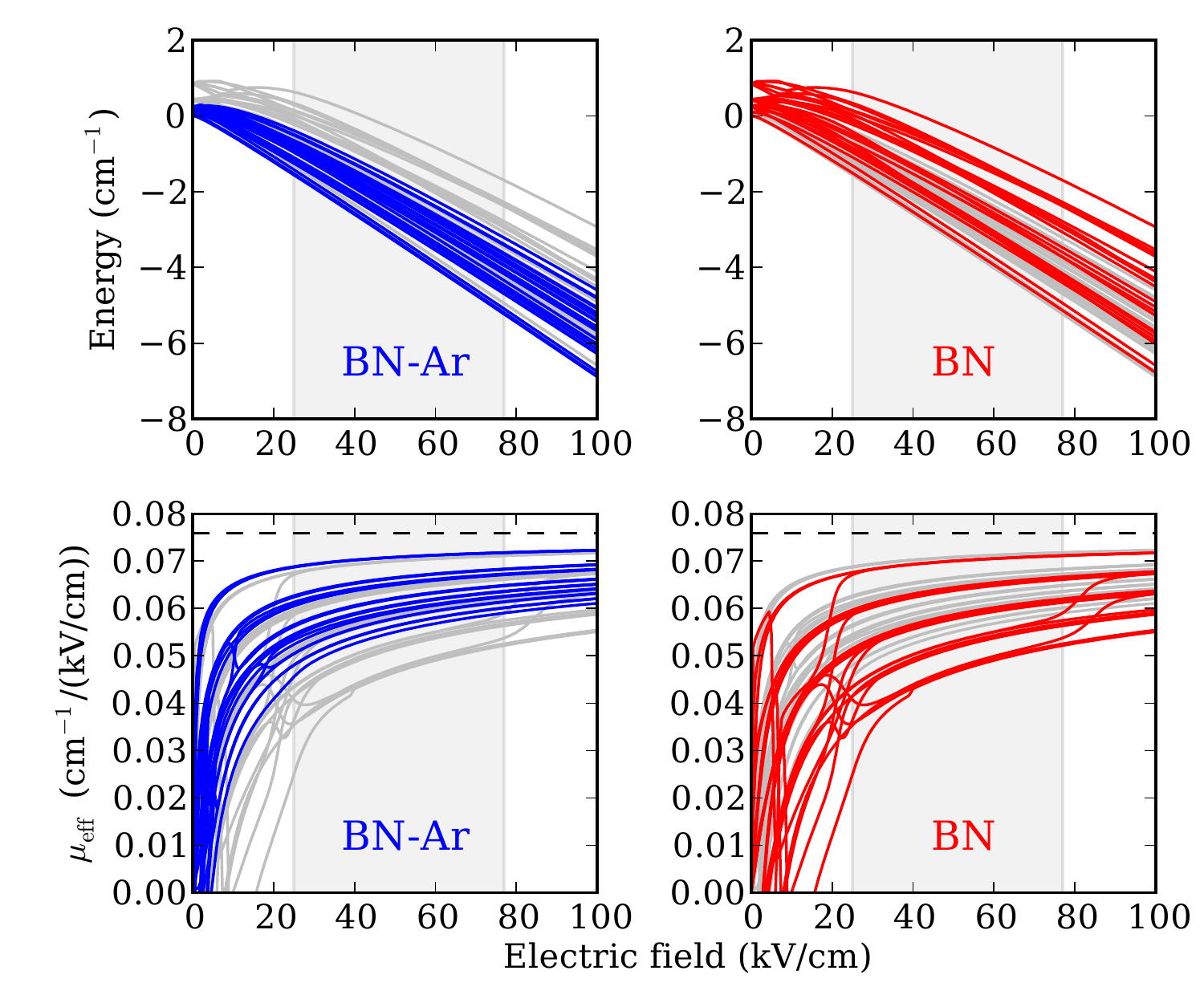}
   \caption{Upper row: Calculated Stark energies of the lowest rotational states ($J\leqslant2$) of the complex (blue) and benzonitrile (red) as function of the electric field. The light grey curves depict the species shown in the respective other column for comparison. Lower row: Calculated effective dipole moments, plotted using the same color scheme as in the upper row. Dashed lines: permanent electric dipole moment of benzonitrile, 4.52~D=0.0759~cm$^{-1}$/(kV/cm). The electric field range sampled by the molecules inside the selector is depicted by the areas shaded in grey.}
   \label{fig:results:starkenergies}
\end{figure}

In \autoref{fig:results:starkenergies}, the Stark energies and effective dipole moments calculated above are depicted for all rotational states of benzonitrile and the complex with $J\leqslant2$. The upper (lower) row shows the energy (effective dipole moment) as a function of electric field strength. Red curves are used for benzonitrile and blue curves for the complex. The grey curves in all four panels depict the species shown in the other column, and the dashed horizontal lines in the lower row show the permanent dipole moment of benzonitrile. Calculations show that at the relevant field strengths -- above 20~kV/cm -- the Stark energies of the complex and its density of states are larger on average than for benzonitrile. This effect is due to the smaller rotational constants of the complex and results in effective dipole moments that, in going to higher field strengths, converge more rapidly to the permanent dipole moment than for benzonitrile.

\begin{figure}
   \centering
   \includegraphics[width=\linewidth]{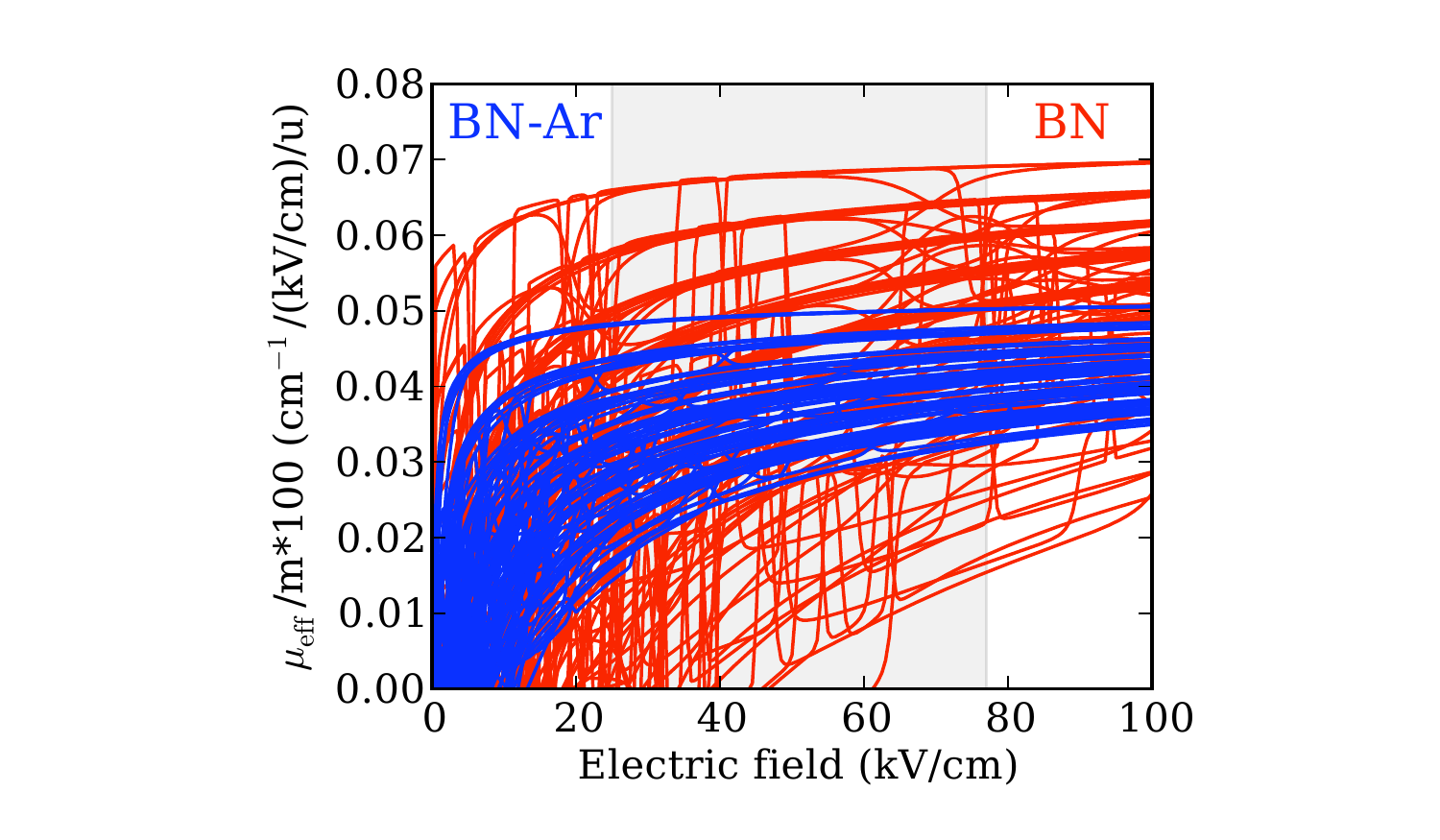}
   \caption{Calculated \textmu$_\textrm{eff}$/m-ratios of the lowest rotational states ($J\leqslant4$) of benzonitrile (red) and the complex (blue). The electric field range sampled by the molecules inside the selector is depicted by the grey shaded area.}
   \label{fig:results:mubym}
\end{figure}

In a molecular beam that consists only of benzonitrile and the complex, both at a rotational temperature below 20~mK, only the ground states of the two species are populated. In this case, transmission through the selector can be restricted to the complex alone, when an ac frequency is chosen below the onset of the benzonitrile transmission curve in \autoref{fig:results:transmissioncurves_d05}, \eg, at 2.7~kHz. Alternatively, a pure benzonitrile beam is obtained by suppressing the transmission of the complex at a 4.5~kHz or higher switching frequency.

However, the typical rotational temperature in a supersonic molecular beam is on the order of 1~K at which the population of large molecules is distributed over many rotational levels. Therefore, the fraction of ground-state molecules is relatively low. As shown in \autoref{fig:results:starkenergies}, the effective dipole moments \textmu$_\textrm{eff}$ of higher rotational levels are equal to or smaller than for the ground state inside the electric fields of the selector. In order to achieve the same focusing for a smaller \textmu$_\textrm{eff}$, the time spent in each of the two field configurations must be longer, \ie, the switching frequency must be lower. This is the rationale for peak A in \autoref{fig:results:guidingspectrum} becoming stronger relative to peak B at low ac frequencies. As a result, the transmission curve for a rotational level either coincides with the ground-state curve or is shifted to lower switching frequencies. This implies that also in a thermal beam transmission of the complex can be suppressed by choosing a 4.5~kHz switching frequency.

Selective suppression of the monomer becomes impossible at such rotational temperatures using only the selector. The transmission through the device is sensitive to the \textmu/m-ratio alone and two molecular species with the same \textmu/m-ratio cannot be distinguished. This is illustrated in \autoref{fig:results:mubym} where the calculated effective dipole moments of benzonitrile and the complex for all rotational states with $J\leqslant4$ are divided by the respective masses. The region above 5$\times10^{-4}$~cm$^{-1}$/(kV/cm)/u is solely populated by rotational states of benzonitrile. By choosing a switching frequency that is only suitable for the \textmu$_\textrm{eff}$/m-ratios in that region, transmission of the complex can be suppressed.

For the complex, however, the curves of all states overlap with the monomer, also for states with higher $J$-values than shown in \autoref{fig:results:mubym}, and, therefore, the selector cannot be operated such that the transmission of the complex is favored while the transmission of benzonitrile is suppressed.

\section{Conclusions}
\label{sec:conclusions}

In this work, we demonstrate the focusing of the benzonitrile-argon van-der-Waals complex in a cold molecular beam using ac electric fields. At 2.8~kHz ac frequency, the experimental excitation spectrum is described by simulations based on a thermal distribution of the initial states at 0.8~K. The transmission characteristics, \ie, the signal intensity as function of ac switching frequency, are described by trajectory simulations in combination with Stark energy calculations. Simulations show that the permanent electric dipole moment of benzonitrile is retained upon the attachment of the argon atom to within $\pm5\%$. By choosing appropriate ac frequencies, transmission of the complex through our selector can be selectively suppressed. When only the ground states of the two species are populated, \eg, near 0~K rotational temperature, transmission can also be restricted to the complex.

\section{Acknowledgments}
\label{sec:acknowledgments}

The authors would like to thank Robert W.~Field and Janneke Blokland for helpful discussions. The design of the electronics by G. Heyne, V. Platschkowski, and P. Schlecht has been crucial for this work. We acknowledge software support by U. Hoppe and H. Junkes.

\bibliography{}
\end{document}